\documentclass[letterpaper]{article}
\usepackage{natbib,alifexi}
\newcommand{\be}{\begin{eqnarray}}
\newcommand{\ee}{\end{eqnarray}}

\title{Darwinian Evolution of Cooperation via Punishment in the ``Public Goods" Game}
\author{Arend Hintze$^{1}$, Christoph Adami$^{1}$ \\
\mbox{}\\
$^1$Keck Graduate Institute, 535 Watson Dr., Claremont CA 91711\\
adami@kgi.edu}

\begin{document}
\maketitle
     \begin{abstract}
The evolution of cooperation has been a perennial problem for evolutionary biology because cooperation is undermined by selfish cheaters (or ``free riders") that profit from cooperators but do not invest any resources themselves. In a purely ``selfish" view of evolution, those cheaters should be favored. Evolutionary game theory has been able to show that under certain conditions, cooperation nonetheless evolves stably. One of these scenarios utilizes the power of punishment to suppress free riders, but only if players interact in a structured population where cooperators are likely to be surrounded by other cooperators. Here we show that cooperation via punishment can evolve even in well-mixed populations that play the ``public goods" game, if the synergy effect of cooperation is high enough. As the synergy is increased, populations transition from defection to cooperation in a manner reminiscent of a phase transition. If punishment is turned off, the critical synergy is significantly higher, illustrating that (as shown before) punishment aids in establishing cooperation. We also show that the critical point depends on the mutation rate so that higher mutation rates discourage cooperation, as has been observed before in the Prisoner's Dilemma.

\end{abstract}

\section{Introduction}
"Tragedy of the commons" is the name given to a social dilemma~\citep{Hardin1968} that occurs when a number of individuals maximize their self-interest by exploiting a public good, and by doing so harm their (and others') own long-term interest. This is but one dilemma~\citep{Frank2006} that can be described within the framework of Evolutionary Game theory~\citep{MaynardSmith1982,Axelrod1984,Dugatkin1997,HofbauerSigmund1998,Nowak2006}. While the tragedy of the commons is important in social science and politics (overfishing, and the destruction of the environment in general come to mind), it also plays an important in role in biology: both the evolution of virulence~\citep{Frank1996} and the manipulation of a host by a group of parasites~\citep{Brown1999} can be viewed as a dilemma of the public goods type.

The public goods game is a standard of experimental economics~\citep{DavisHolt1993,Ledyard1995}, where players possess a number of tokens that they can contribute to a common pool (the ``investment" into the public good). The total contributed by the players is multiplied by a ``synergy factor", and this amount is then equally distributed to the players in the pool, irrespective of whether they have contributed or not. A group of players fares best if all the players contribute so as to take maximum advantage of the synergy, but this behavior is vulnerable to ``free-riders" that share in the pool but do not contribute themselves. In fact, the rational Nash equilibrium of the game is for all players to withhold their tokens. 

It has been shown that {\em punishment} is an effective way to counteract defectors~\citep{FehrGachter2002,FehrFischbacher2003,Hammerstein2003,NakamaruIwasa2006,CamererFehr2006,Gurerketal2006,Sigmundetal2001,HenrichBoyd2001, Boydetal2003,Brandtetal2003,Helbingetal2010}. Because punishment involves an additional cost to the co-operators that already invest into the public good~\citep{Yamagishi1986,Fehr2004, Colman2006}, these cooperators (termed ``moralists" by Helbing et al.~\citeyear{Helbingetal2010}) are themselves vulnerable to the invasion of non-punishing cooperators called ``secondary free-riders". As a consequence, we might expect that moralists ultimately become extinct, either because they were outcompeted by defectors, or by cooperating free-riders who benefit from the punishment without the associated cost. Alternatively, if moralists are ultimately successful in eliminating defectors, the punishment gene stops to be under selection and should drift, again resulting in the demise of moralists. 

It has recently been shown that, instead, in simple spatial games, moralist can win direct competitions~\citep{Helbingetal2010} if the environmental conditions are favorable, namely if the cost to benefit ratio of punishment favors moralists over defectors. Spatial games, where the offspring of successful strategies are placed near the parent, and where as a consequence strategies are more likely to play against kin strategies, give rise to spatial reciprocity~\citep{Sigmundetal2001}. This appears to be the advantage that moralists need to gain superiority. In the simulations of Helbing et al., evolution proceeded by the {\em imitation} of successful neighboring strategies rather than Darwinian evolution, but the dynamics are similar. However, because strategies in those simulations are deterministic (limiting genetic space to four genotypes), large grids had to be used in order to prevent premature extinctions.

Here, we show that spatial reciprocity is in fact not a necessary condition for the evolution of cooperation via punishment and the dominance of moralists, if stochastic strategies can evolve via Darwinian dynamics in a framework where decisions are encoded within genes that adapt to their environment. There are conditions where cooperation evolves even without punishment, but absent those, punishment can promote the evolution of cooperation, as long as punishment is effective and cheap, in well-mixed populations. If cooperation becomes so dominant that defectors are brought to extinction, the punishment gene drifts to neutrality. Finally, we also observe that stable environments that are believed to be more predictable for players also increase the chance for cooperators to evolve and to be stable, as observed earlier within the iterated Prisoner's Dilemma~\citep{Iliopoulosetal2010}.

\section{Experimental Design}
We evolve stochastic strategies playing the public goods game with punishment. Each individual in a group of $k$ players ($k=5$ in the present implementation) can decide to cooperate by making a contribution of 1 unit to the public good, while defecting individuals do not contribute. We encode this choice as a probability $p_C$, which can be thought of as the outcome of a network of genes that encode this decision. When mutating strategies, instead of mutating the individual genes that make up the decision pathway, we simply replace the parental probability $p_C$ by a uniformly drawn random number in the offspring. We will call the locus encoding the probability $p_C$ simply the ``C gene". 

The sum of all contributions from cooperating players is multiplied by $r$ (the synergy factor) and divided among all players. In addition, each player has the option to punish players who do not contribute. This decision is encoded by an independent probability $p_P$, called the ``P gene". Following Helbing et al.~\citeyear{Helbingetal2010}, those players who defect suffer a fine $\beta/k$ levied by the punishers in the group, whereas the punishers incur a penalty of $\gamma/k$. At each update, every player engages in a game with all its assigned opponents. The number of cooperators $N_C$, defectors $N_D$, moralists $N_M$ and immoralists (players who defect but also punish~\cite{Helbingetal2010}) $N_I$ is computed, and the payoff is assigned as follows: A cooperator receives 
\be
P_C = r\frac{(N_C+N_M+1)}{k+1} -1\;, \label{eq1}
\ee
while a defector takes away
\be
P_D=r\frac{(N_C+N_M)}{k+1}-\beta\frac{(N_M+N_I)}{k}\;. \label{eq2}
\ee
Moralists  receive 
\be 
P_M=P_C-\gamma\frac{(N_D+N_I)}{k}\;,
\ee
while immoralists earn
\be
P_I=P_D-\gamma\frac{(N_D+N_I)}{k}\;.
\ee

The population consists of 1,024 individuals who each have four assigned opponents. Since all opponents are also players, each individual plays five games per update. The choices of each individual are determined by their probabilities to cooperate $p_C$ and to punish $p_P$. After each round, 2 percent of the population is replaced using a Moran-process~\citep{Moran1962} in a well-mixed fashion, that is, the identity of the players in the group is unrelated to their ancestry so that,  effectively, the members of a particular playing group are randomly selected from the population. We verified that the probability for a player to encounter cooperators is independent of whether that player is a cooperator or a defector, as is required for well-mixed populations~\citep{FletcherDoebeli2009}.  Players that are not replaced are allowed to accumulate their score, which is used to calculate the probability that this player's strategy will be chosen to replicate and fill the spot of a player that was removed in the Moran process.

Every individual's genes mutates with a probability $\mu$ when replicated. As mentioned earlier, the mutation of a gene replaces the probability with a uniformly distributed random number. After 500,000 updates, the line of descent (LOD) of the population is reconstructed, by picking a random organism of the final population and following its ancestry all the way back to the starting organism, which has $p_C=0.5$ and $p_P=0.5$. Because there is only one species in these populations, the LOD of the population coalesces to a single LOD (which is why it is sufficient to pick a random genotype for following the LOD). 

As the strategies adapt to the environmental conditions (specified by the parameters that define the game, as well as the spatial properties, the mutation rate, and the replacement rate), the probabilities that appear on the LOD tell the story of that adaptation, mutation by mutation. While the LOD in each particular run can show probabilities varying wildly, averaging many such LODs can tell us about the selective pressures the populations face. In particular, averaging the probabilities on the LODs after they have settled down (from the transient beginning at the random strategy $(p_C,p_P)=(0.5,0.5)$) can tell us the {\em fixed point} of evolutionary adaptation~\citep{Iliopoulosetal2010}. We determine this fixed point by discarding the first 250,000 updates of every run (the transient), along with the last 50,000 (in order to remove the dependence of the LOD on the randomly chosen anchor genotype) and averaging the remaining 200,000 updates. Note that this fixed point is a computational fixed point only: we do not mean to imply that the population's genotypes all end up on this exact point. Rather, due to the nature of the game, the evolutionary trajectories approach this point and then fluctuate around or near it. Thus, the fixed point reflects the {\em mean} successful strategy given the conditions of the game. 

\section{Results}
\begin{figure}[t]
\begin{center}
\includegraphics[width=3in,angle=0]{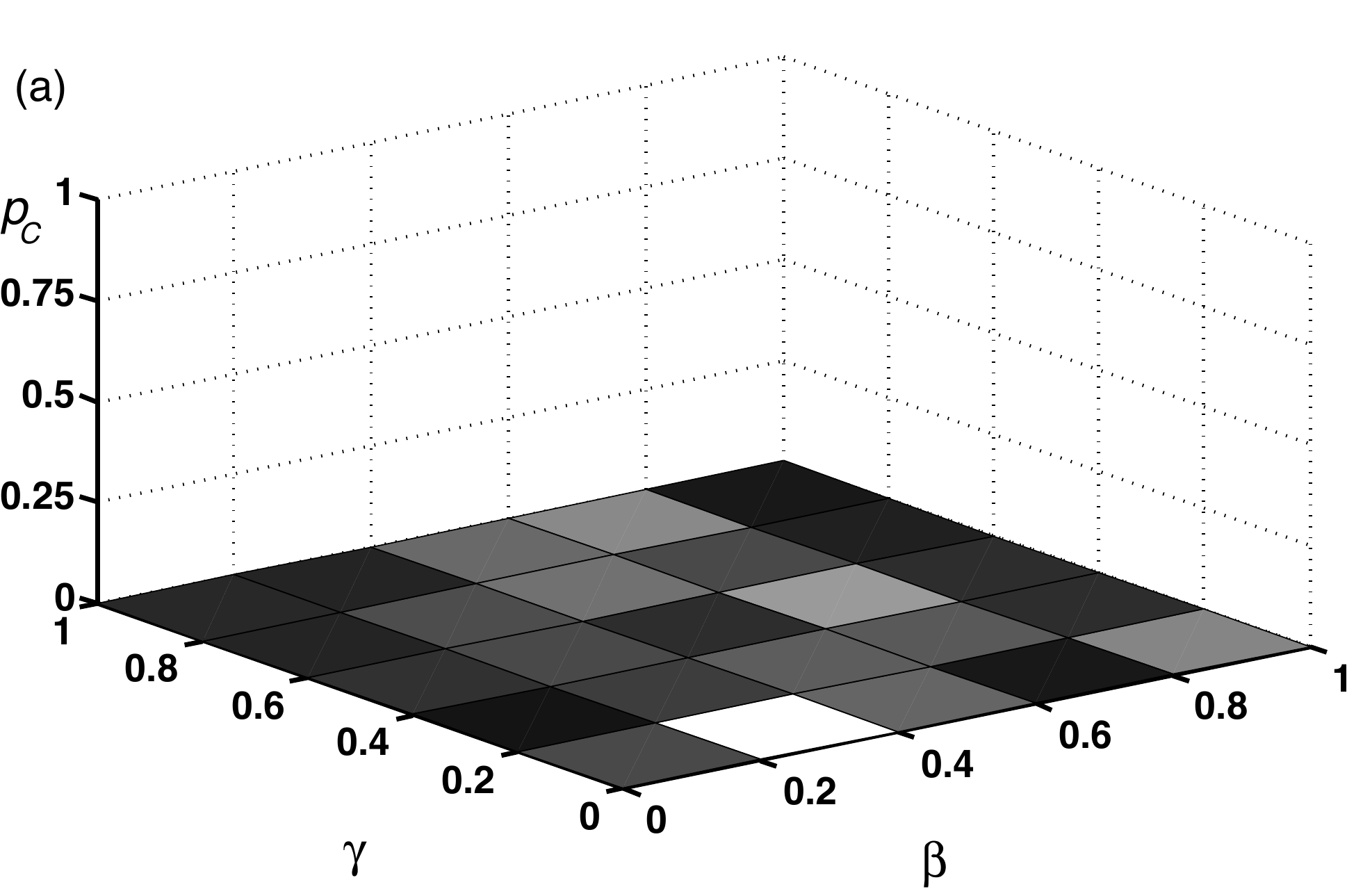}\\
\includegraphics[width=3in,angle=0]{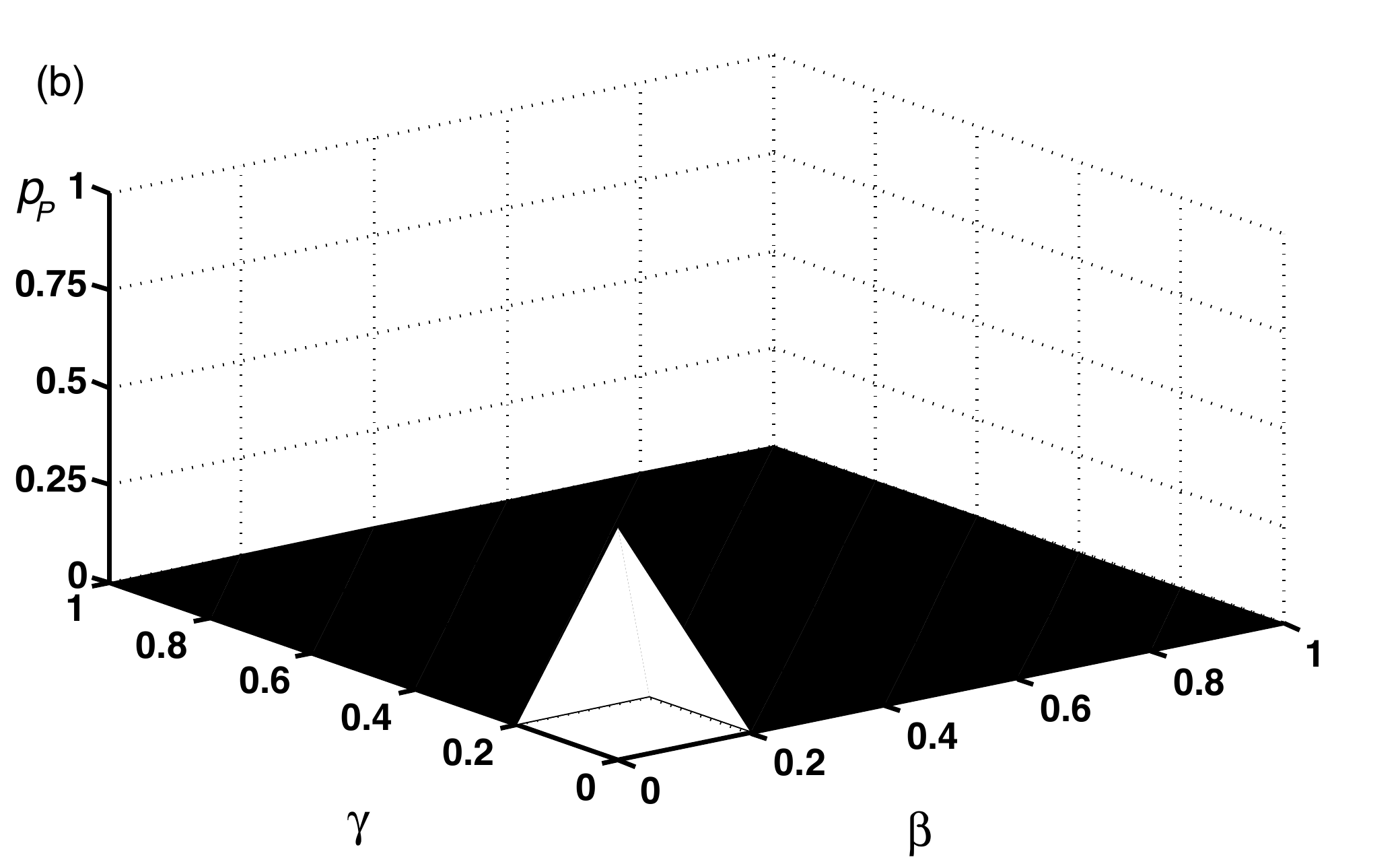}
\caption{(a) Mean probabilities for cooperation $p_C$ at the evolutionary fixed point. Despite the different grey scales, the probability to cooperate vanishes except for noise. (b) The probability to punish $p_P$ at the average fixed. For both genes, $\beta$ and $\gamma$ range from 0.0 to 1.0 in increments of $p_C$, at a fixed $r=3$. We used a mutation rate $\mu=0.02$ per gene in this and the following three figures. Here, we averaged two replicates for each set of parameters.}
\label{fig-r3}
\end{center}
\end{figure}

When mapping the possible parameters $\beta$ (effectiveness) and $\gamma$ (cost) each in the range from 0.0 to 1.0 and at low synergy $r=3.0$, we find that defection is the most prevalent strategy on the LOD (see Figures~\ref{fig-r3}a and b), as was found previously~\citep{Brandtetal2003,Helbingetal2010}. When $\beta$ and $\gamma$ vanish, punishment has no effect,  nor is there a cost associated with that punishment. At this point, the P gene is not under selection and drifts. A drifting gene can be recognized by a mean of 0.5 and a variance of $1/12\approx 0.083$ at the fixed point, as expected for the average and variance of a uniform random number on the interval (0,1). Thus, for this value of synergy (and lower), we find that the strategy fixed point is defection without punishment, except for the values $\gamma$=$\beta$=0, where punishment is random.

\begin{figure}[t]
\begin{center}
\includegraphics[width=3in,angle=0]{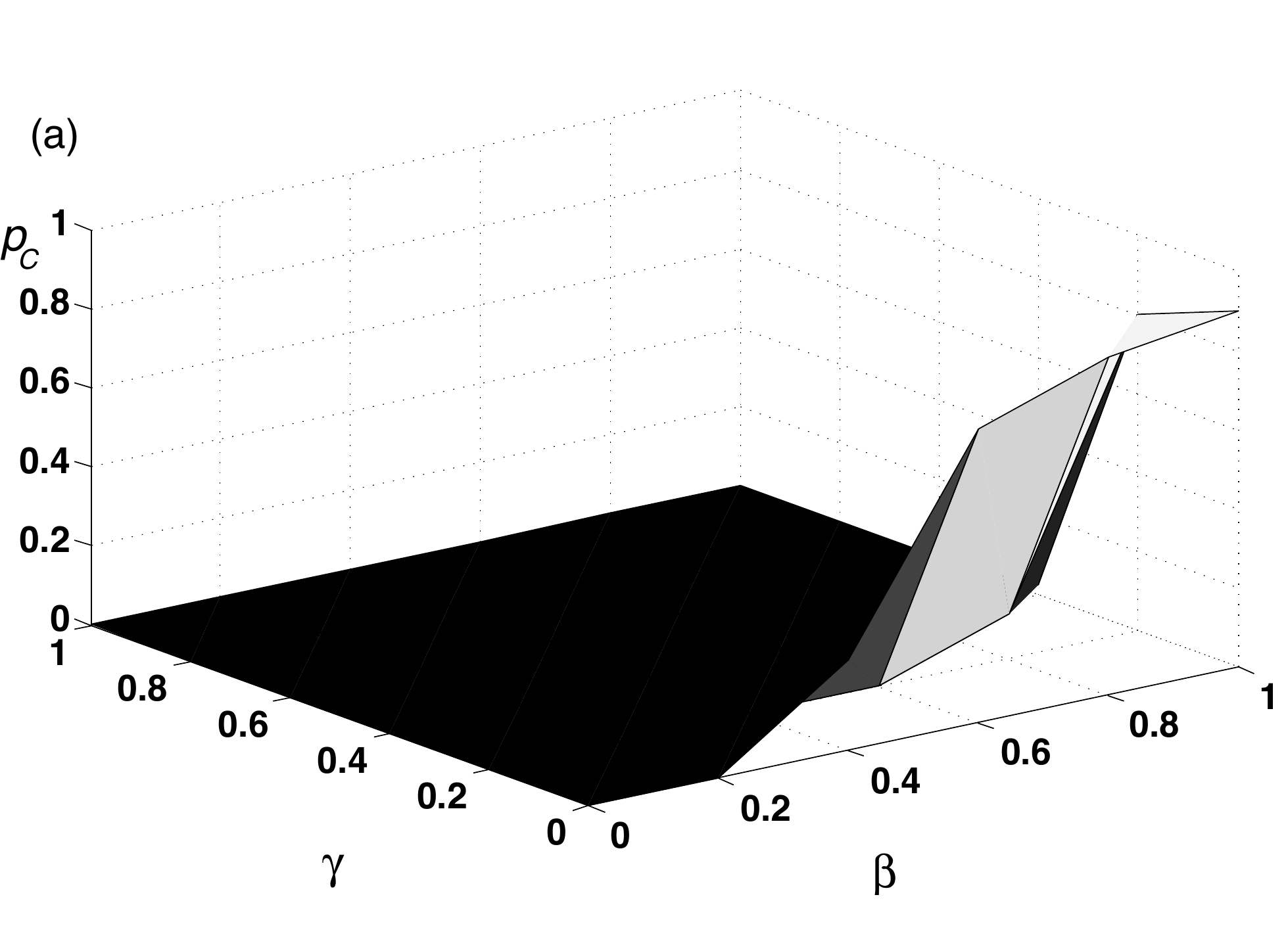}\\
\includegraphics[width=3in,angle=0]{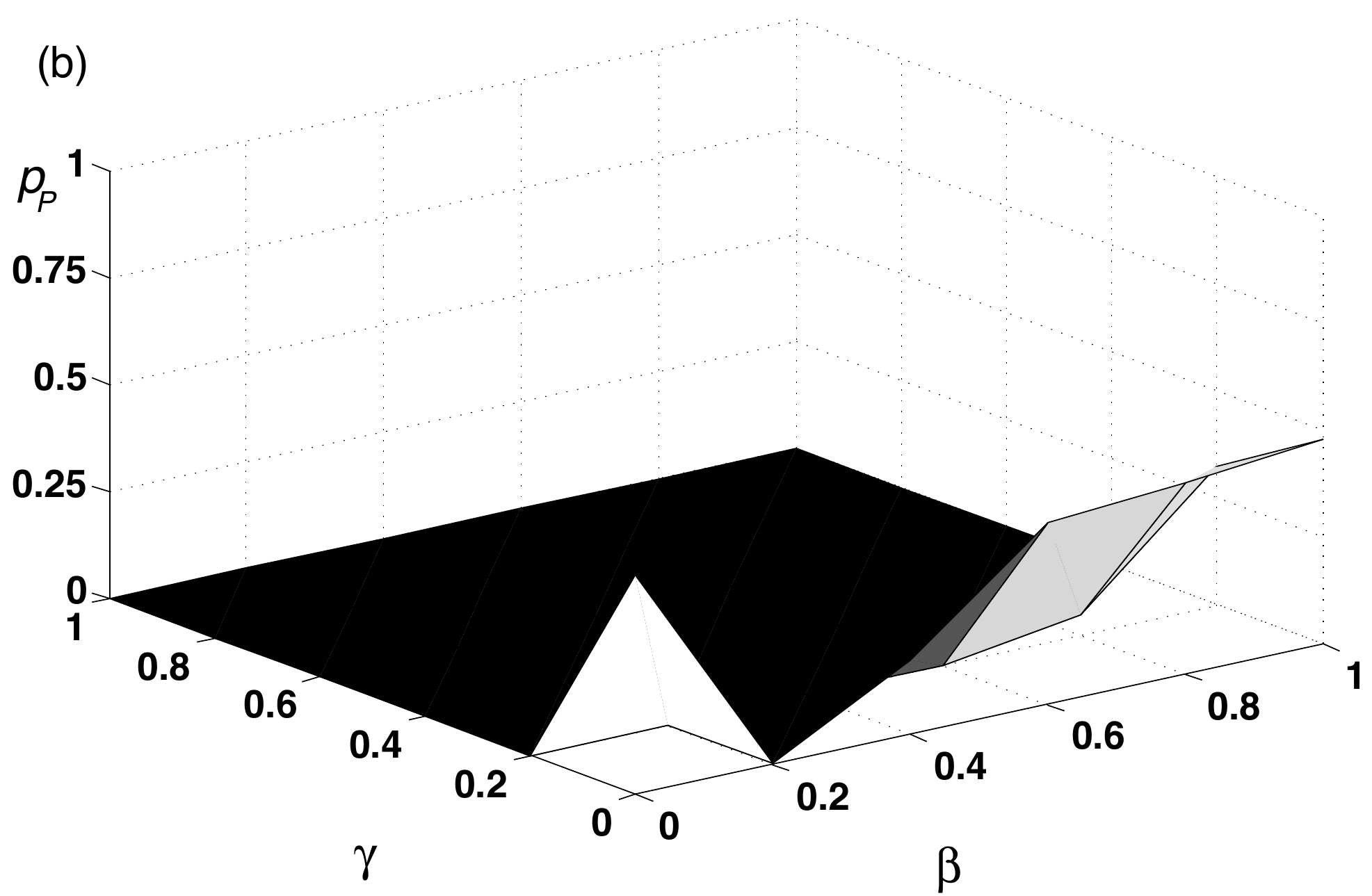}
\caption{Mean probabilities for $p_C$ (a) and $p_P$ (b) averaged over the latter part of the LOD (average fixed point), for $\beta$ and $\gamma$ ranging from 0.0 to 1.0, in increments of 0.2, at $r$=4 (5 replicates per data point).}
\label{fig-r4}
\end{center}
\end{figure}

As the degree of synergy increases to $r=4$, cooperation starts to appear even in this well-mixed population (while it appears as early as $r=2$ for sufficiently high $\beta$ and low $\gamma$ in the spatial version of the game, see~\citealp{Brandtetal2003,Helbingetal2010}). We find players cooperating ($p_C\approx0.8$) at high $\beta$ and low $\gamma$ (see Figure~\ref{fig-r4}a), which indicates that under conditions where punishment is not very costly or even free, punishment pays off. In addition we notice that the probability to punish increases under the same conditions that allows cooperation (high $\beta$ and low $\gamma$, that is high impact, low cost of punishment), indicating that punishment is indeed used to enforce cooperation (Fig.~\ref{fig-r4}b). The mean punishment probability grows to 0.5, but at the same time the variance shows that this gene is not under drift (data not shown). Still, the distribution of probabilities on the LOD is fairly broad, indicating that periods of strong punishment give way to periods where agents are much more forgiving. 
Thus, it appears that punishment under these conditions is effective even if it is engaged in only intermittently.

\begin{figure}[t]
\begin{center}
\includegraphics[width=3in,angle=0]{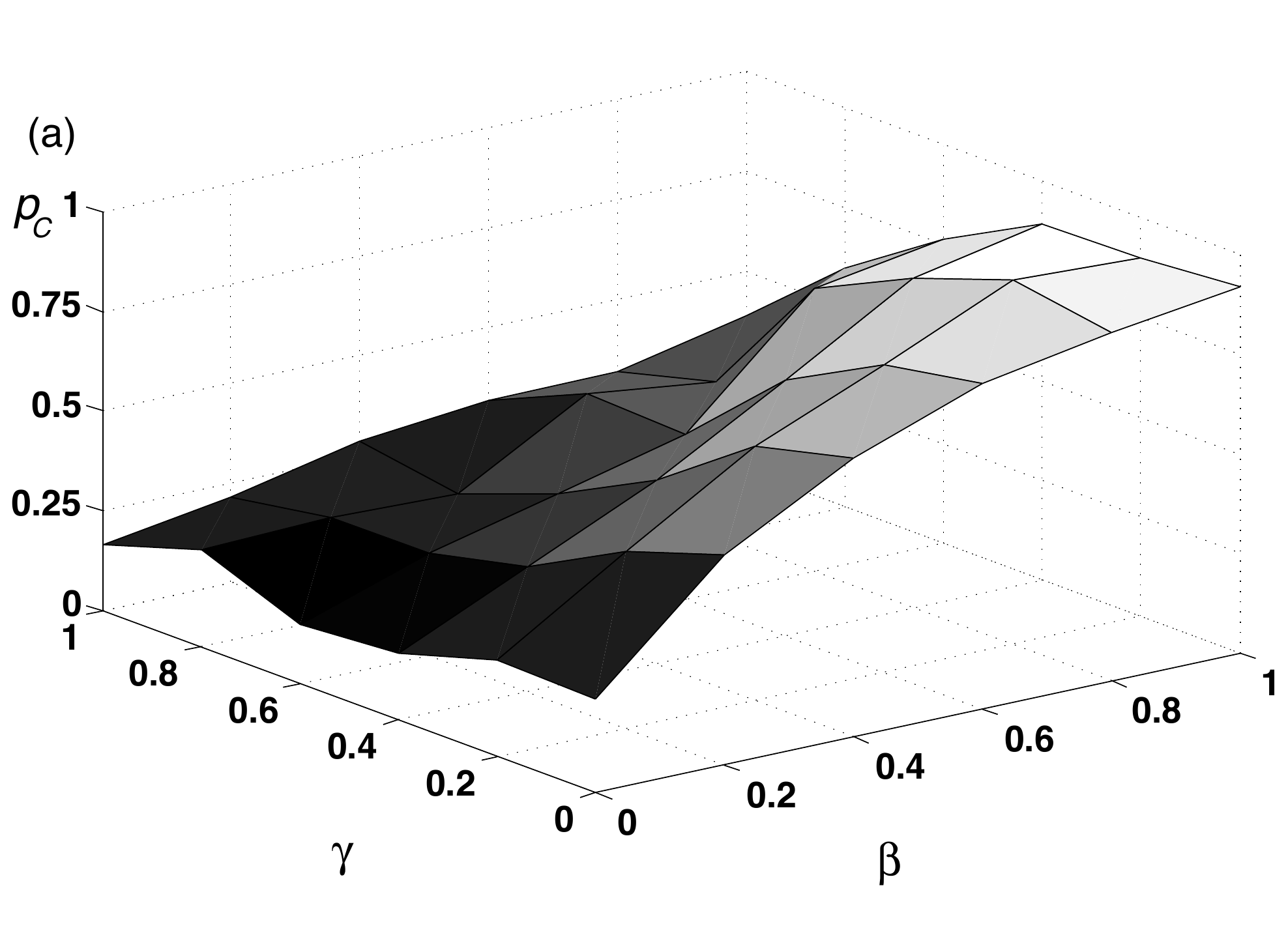}\\
\includegraphics[width=3in,angle=0]{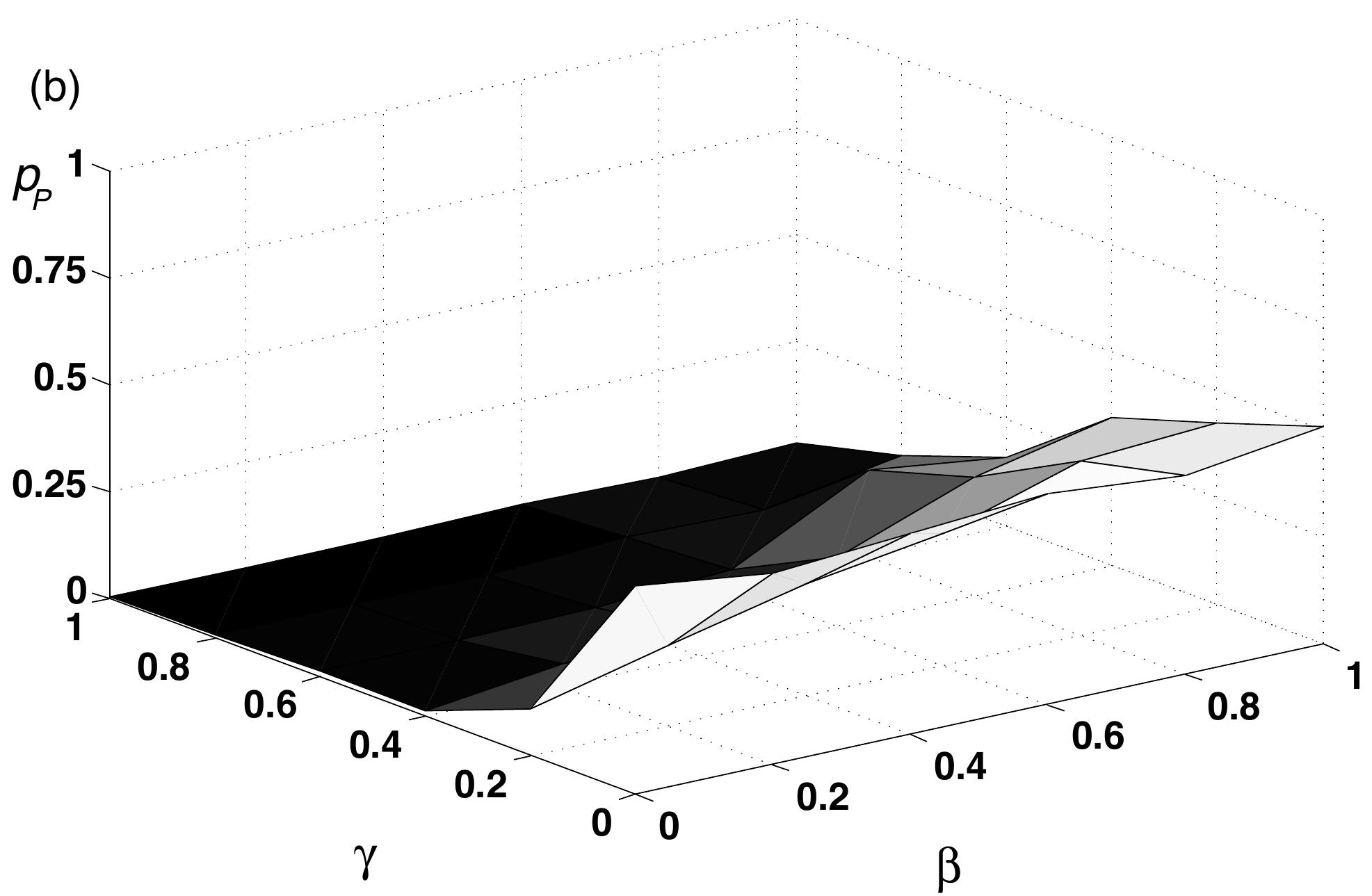}
\caption{Mean probabilities for $p_C$ (a) and $p_P$ (b) at the evolutionary fixed point, for $\beta$ and $\gamma$ ranging from 0.0 to 1.0 in increments of 0.2, at $r$=4.5. 15 replicates per data point.}
\label{fig-r45}
\end{center}
\end{figure}

\begin{figure}[t]
\begin{center}
\includegraphics[width=3in,angle=0]{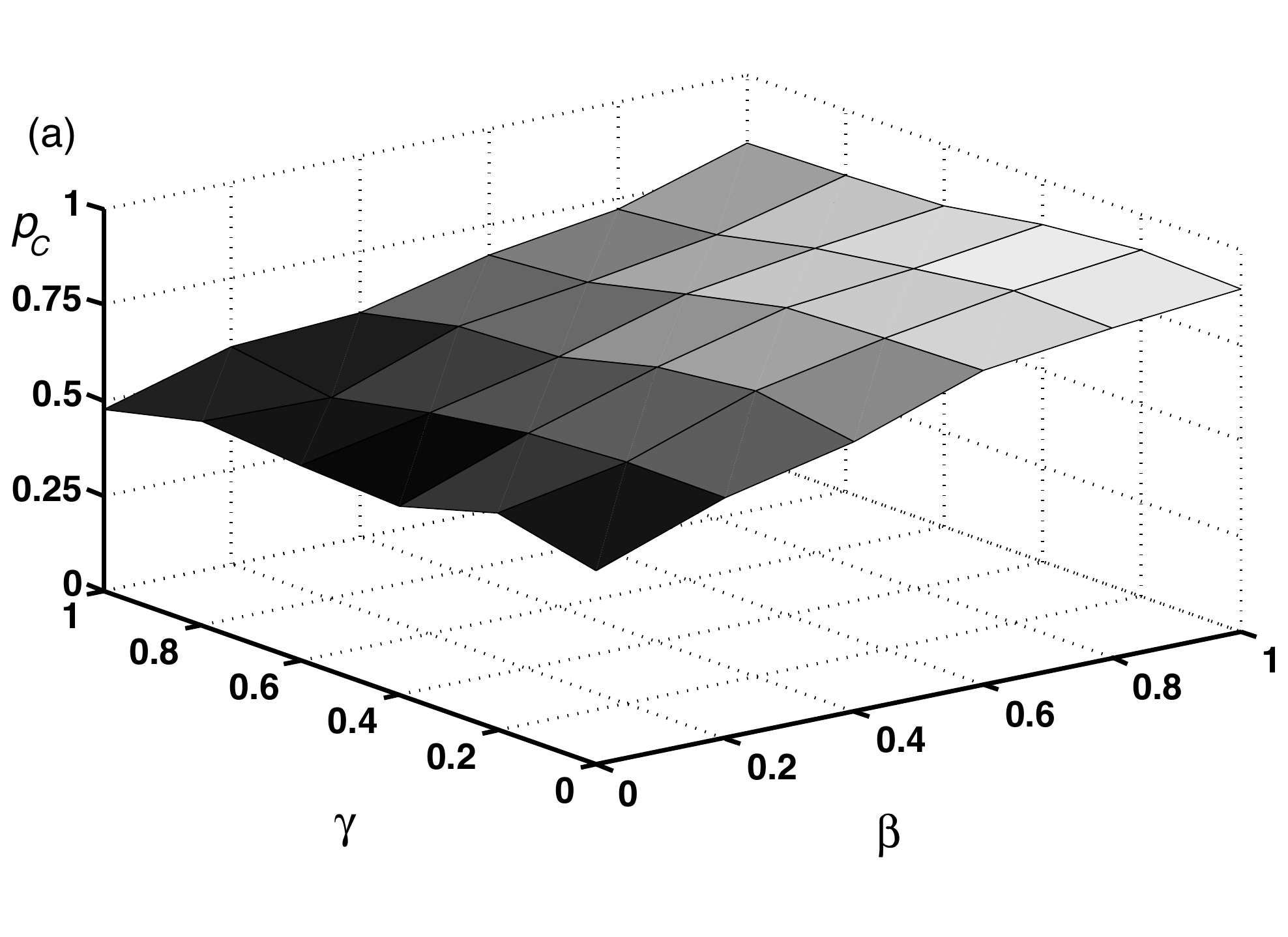}\\
\includegraphics[width=3in,angle=0]{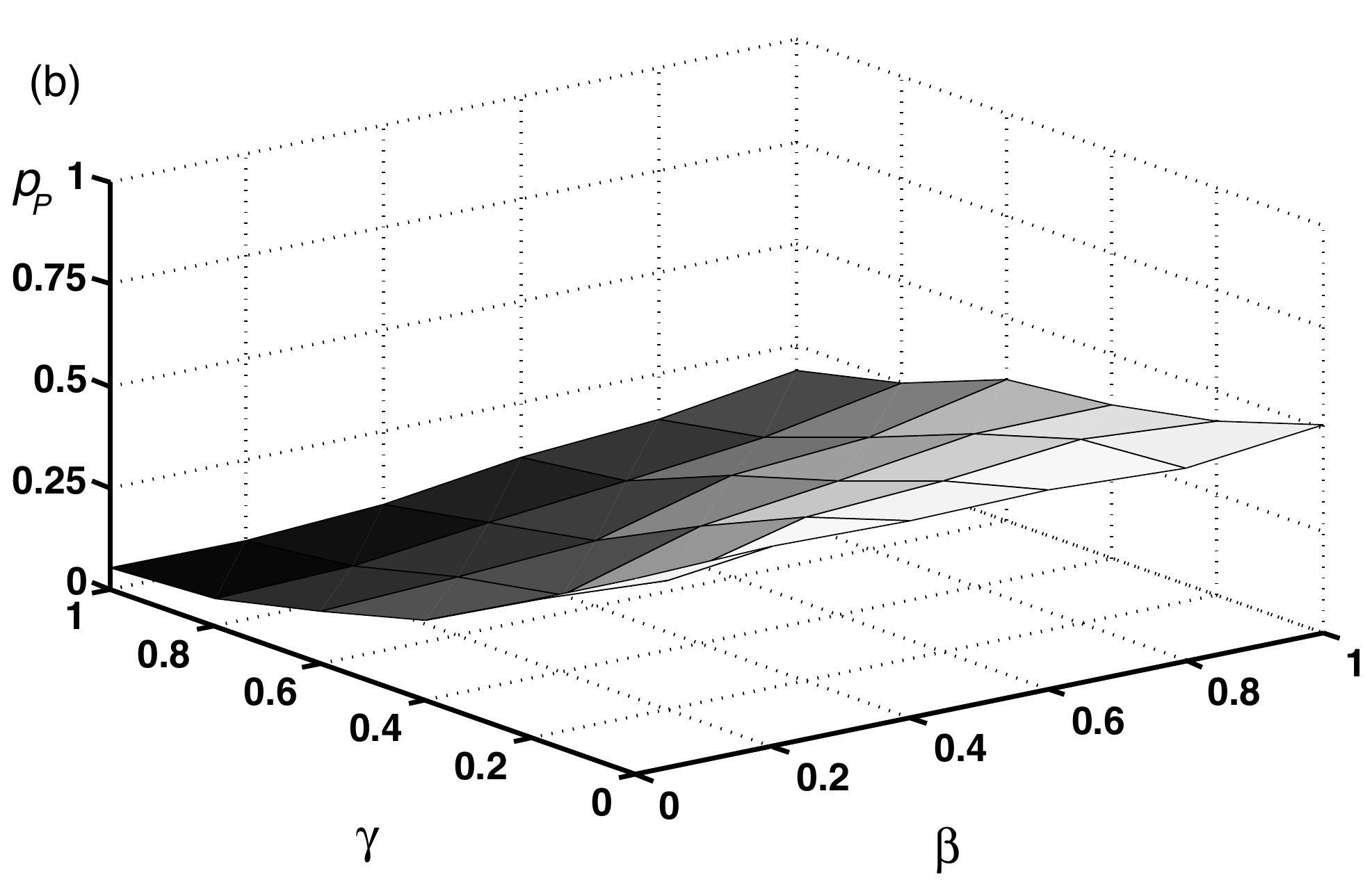}
\caption{Mean probabilities for $p_C$ (a) and $p_P$ (b) at the evolutionary fixed point, for $\beta$ and $\gamma$ ranging from 0.0 to 1.0 in increments of 0.2, at $r$=5 (five replicates per data point).}
\label{fig-r5}
\end{center}
\end{figure}

Increasing the synergy level even higher towards $r$=4.5 shows the emergence of dominance of cooperation ($p_C>$0.5) for most of the range of punishment cost and effectiveness, see Figure~\ref{fig-r45}a. At the same time the punishment probability reaches 0.5 for a larger range of parameters (Fig.~\ref{fig-r45}b), but the mean punishment probability on the LOD never exceeds 0.5, implying that full persistent punishment is not stable, and probably not necessary. Increasing synergy to $r=5$ reveals a population that engages in cooperation for almost all parameter settings (see Figure \ref{fig-r5}), even at conditions where punishment is costly without much impact ($\beta<0.5$, $\gamma>0.5$) but the variance suggests that at high punishment effect and low cost,  this gene may be drifting (as it is only selected for if defectors are prominent). This outcome is expected because at $r=5$, the cooperators' payoff is equal to or higher than the defectors, and exactly equal in the absence of punishment (see below). Thus, defectors should disappear and punishment become random. 

Note that, in an implementation where decisions are deterministic (such as in the implementation of Helbing et al.,2010), punishment may remain for a long time in the population even though it is not selected anymore. In fact, Helbing et al. (2010) increased their population sizes precisely because they observed the disappearance of punishment in smaller cooperating population. From what we have observed here, this disappearance is due entirely to neutral drift.


\section{Critical Behavior}
\begin{figure}[t]
\begin{center}
\includegraphics[width=3.1in,angle=0]{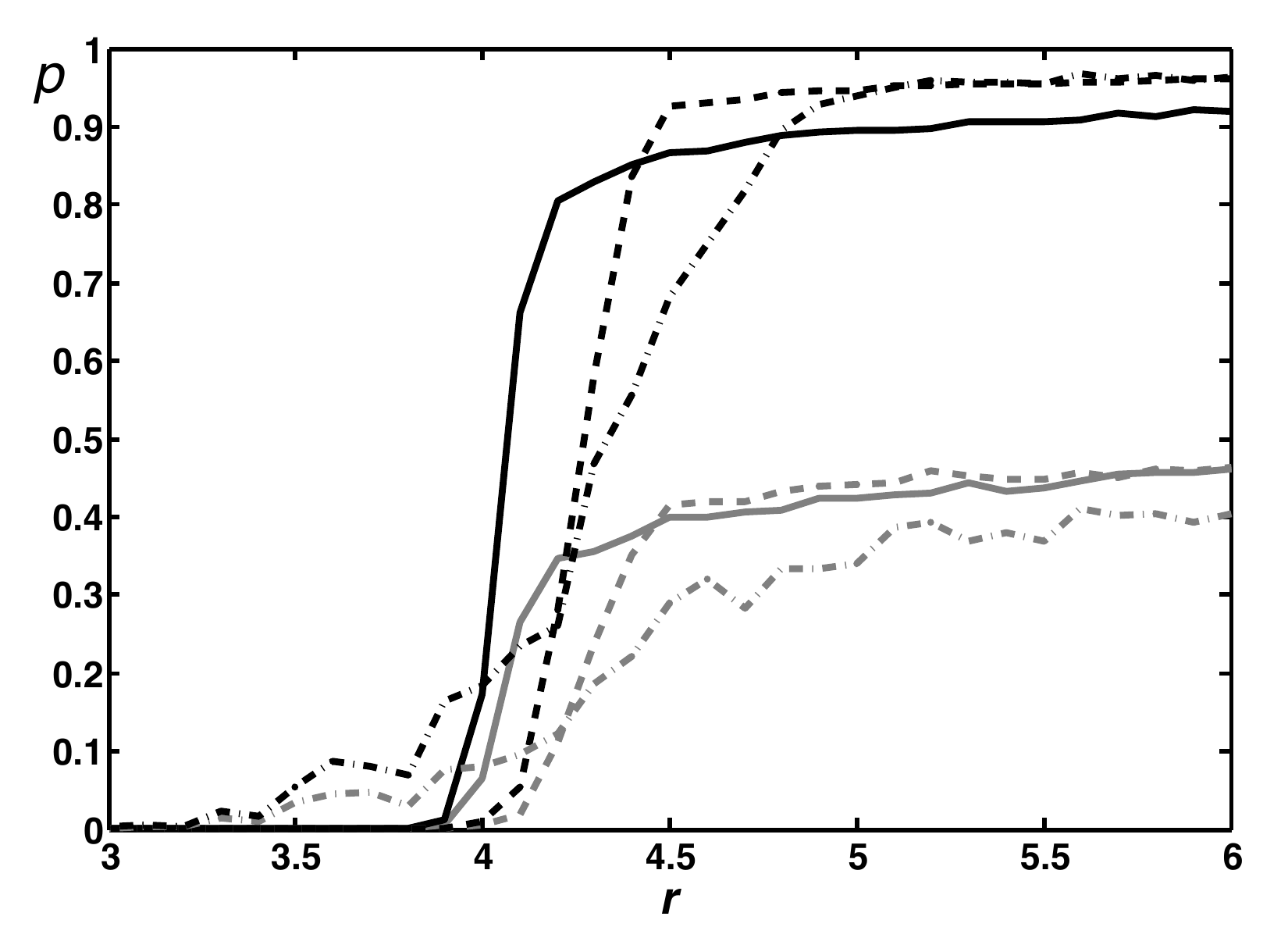}
\caption{Mean probability of cooperation $p_C$ (black lines) and punishment $p_P$ (grey lines) at the evolutionary fixed point of the trajectory, as a function of the synergy $r$ for three different mutation rates: dash-dotted: $\mu=0.001$, dashed: $\mu=0.01$, and solid: $\mu=0.02$ (100 replicates for each data point). }
\label{fig-mut}
\end{center}
\end{figure}

Previously, a phase transition between cooperative and defective behaviour in the public goods game was observed for the spatial version~\citep{SzaboHauert2002,Brandtetal2003} of the game (but not the well-mixed version).
In Fig.~\ref{fig-mut} we show the mean probability at the evolutionary fixed point of both the C gene (black lines) and the P gene (grey lines) as a function of the synergy level $r$, for different mutation rates (dash-dotted lines: $\mu=0.001$, dashed lines: $\mu=0.01$ and solid lines: $\mu=0.02$, which is the mutation rate we used in Figs~1-4). We note the sudden emergence of cooperation at a critical synergy level, but that this level depends on the mutation rate. For the highest mutation rate (black solid line in Fig.~\ref{fig-mut}) cooperation emerges the earliest, however the critical point (defined as the point where the cooperation probability reaches 0.5) moves towards higher synergy levels.  As the mutation rate is lowered, the critical point moves to the left and the fixed point probability is higher. The emergence of punishment (grey lines in Fig.~\ref{fig-mut}) follows the same trend, and again we notice that the mean never exceeds 0.5. Thus, we see that higher mutation rates lead to higher critical synergy values necessary to enable cooperation, in other words, the more uncertain environments are discouraging for cooperators as observed earlier~\citep{Iliopoulosetal2010}.

\begin{figure}[t]
\begin{center}
\includegraphics[width=3.3in,angle=0]{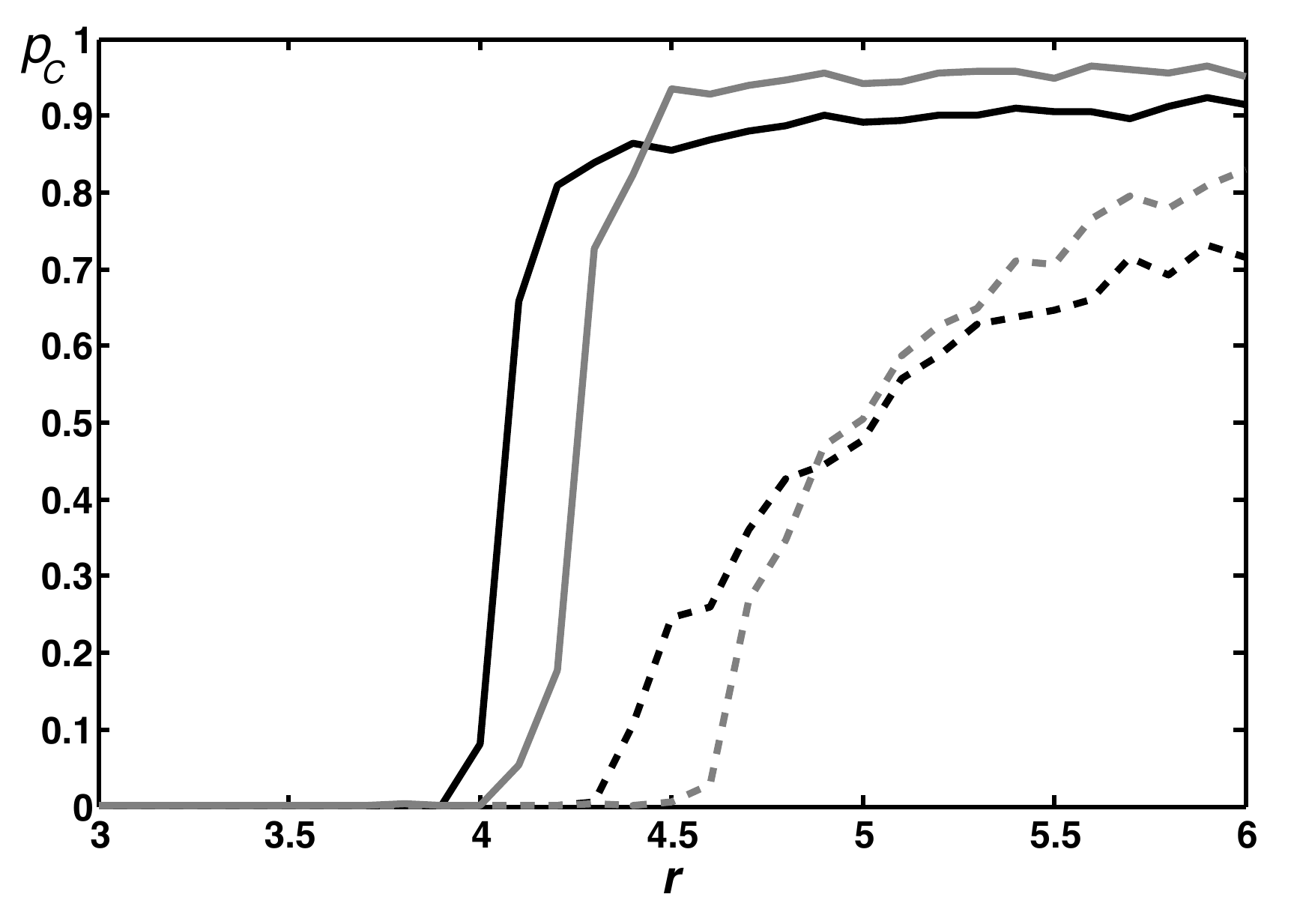}
\caption{Mean probabilities for $p_C$ at the fixed point, for effectiveness $\beta=0.8$ and cost of punishment $\gamma=0.2$, as a function of synergy  $r$. Solid lines are the standard protocol, while the dashed lines represent experiments with punishment turned off ($p_P=0$). Black: $\mu=0.02$, grey: $\mu=0.01$, 20 replicates per curve}
\label{fig-trans}
\end{center}
\end{figure}

It is instructive to study how punishment affects the critical point. To do this, we ran a control of the experiment where punishment did not exist. In that case, we observe a critical $r$ that is significantly higher that what we observe with punishment (see Fig.~\ref{fig-trans}), showing again how punishment aids in the establishment of cooperation. Note also that the levels of cooperation achieved are significantly higher when punishment exists.

We can calculate approximately the point at which cooperation is favored in a mean-field approach that does not take mutation and evolution into account,  by writing Eqs.~(\ref{eq1}-\ref{eq2}) in terms of the density of cooperators $\rho_C$ encountered by players in a group.
Both naked cooperators and punishing cooperators (moralists) contribute to this density, i.e., $\rho_C=(N_C+N_M)/N$, where $N$ is the total number of players in the group. We can also introduce the mean density of punishers $\rho_P=(N_M+N_I)/N$ encountered by a player. Because the mean density of cooperators and punishers is the {\em same} for both cooperators and defectors in a well-mixed scenario (but not for spatial play!), we can then write
\be
P_C = r\frac{k\rho_C+1}{k+1} -1\; \label{eq-coop}
\ee
and
\be
P_D=r\frac{k\rho_C}{k+1}-\beta\rho_P \label{eq-def}\;,
\ee
and we expect cooperation to be favored if
\be
P_C-P_D=\frac{r}{k+1}-1+\beta\rho_P>0\;
\ee
or
\be
r>(k+1)(1-\beta\rho_P)\;.\label{crit}
\ee

This equation implies that the emergence of cooperation depends crucially on the density of punishers. In fact, the mean-field theory predicts that cooperation in the absence of punishment emerges only at $r=5$, while we see it emerge quite a bit earlier than that (see Fig.~\ref{fig-trans}, dashed lines). Note, however, that the critical point moves towards the predicted value $r=5$ as the mutation rate is lowered, which would not be surprising as the theory holds strictly only for vanishing mutation rate.
Because we expect that the density of punishers increases as the mutation rate increases (because mutations can introduce defectors at an elevated rate, necessitating a more pronounced punishment response), we can also expect the critical mutation rate to drop commensurately, but it is clear from the previous comment that there are mutation rate effects in the dynamics of the population that are independent of punishment.

Because of the critical importance of punishers in determining the synergy level at which cooperation emerges, the public goods game with a genetic basis implies a curious dynamics close to the critical point. Below the critical point, defection is a stable strategy, and punishment is absent. Only when cooperation emerges as a possibility, punishment becomes more and more important, leading to a lowering of the critical synergy for cooperation via Eq.~(\ref{crit}). Thus, cooperation emerges rapidly and decisively once a critical level has been achieved.
Once cooperation is dominant and defectors are all but driven to extinction, punishment becomes irrelevant and the gene begins to drift. As this happens, the fraction of punishers drops, raising the critical synergy. Thus, a drifting punishment gene can lead to the sudden re-emergence of defectors as stable states. Once those have taken over, the reverse dynamics begins to unfold. In other words, we should observe periods of cooperation and defection follow each other closely as the synergy is near the critical point. 

This dynamics is reminiscent of the phenomenon of supercooling and superheating in phase transitions. If we imagine the synergy parameter $r$ as the critical parameter and the mean probability to cooperate as the order parameter, it is possible that when $r$ is slowly increased, the population remains in the defecting phase because a switch to cooperation requires a critical number of cooperators as a ``seed". In such a situation, the defecting phase is unstable to fluctuations. If a critical number of cooperators emerge by chance, punishment immediately becomes effective, lowers the critical point as implied by Eq.~(\ref{crit}), and the population could transition to cooperation very quickly. The shape of the critical curves in Fig.~\ref{fig-mut} supports this point of view: at higher mutation rates, the transition from defection to cooperation as the synergy $r$ is increased is much more gradual, presumably because the increased mutation rate increases the probability to create the seed of cooperators that is necessary for the emergence of the cooperative phase. An investigation of the population dynamics at the critical point will be the subject of a subsequent investigation.

\section {Discussion}

We studied Darwinian evolution of stochastic strategies in the public goods game for well-mixed populations, using genes that encode the probabilities for cooperation and punishment. It is known that punishment can drive the evolution of cooperation above a critical synergy level as long as there is a spatial structure in the environment~\citep{Brandtetal2003,Helbingetal2010}. It was also previously believed that in well-mixed populations cooperation can only become successful if additional factors like reputation~\citep{Sigmundetal2001} are influencing the evolution. Here we show that cooperation readily emerges in a well-mixed environment above a critical level of synergy. This critical level is influenced by a number of factors, such as the rate of punishment and the mutation rate. 

If the conditions for punishment are good (that is, the cost for punishment is low and the effect is high) we find cooperative strategies that also have elevated probabilities to punish, that is, they are moralists. But if punishment is cheap and effective, we also see that defectors practically vanish, which in turn obviates the need for punishment, so much so that the punishment gene begins to drift. This effect, however, is also mutation rate dependent, because higher mutation rates will automatically create a higher influx of defectors even if they cannot be maintained by selection.

We conclude that in well-mixed populations cooperation can emerge if the synergy outweighs the defectors' reward. If the mutation rate is low enough, the loss of defectors makes punishment obsolete, that is, the selective pressure to punish disappears. Naturally, once this has occurred defectors can again gain a foothold, and the balance of power between cooperators and defectors could shift. Such a shift, however, reinstates the selective pressure to punish, leading to a re-emergence of moralists that can drive defectors out once more. Thus, for synergy factors near the critical point, we can expect oscillations between cooperators and defectors, and no strategy is ever stable~\citep{Hintzeetal2010}.


\section{Acknowledgements}
We  thank the members of the Evolutionary Dynamics group at KGI for discussions.  
This work was supported by the National Science Foundation's Frontiers in Integrative Biological Research grant FIBR-0527023. 
\footnotesize
\bibliographystyle{apalike}
\bibliography{PD}

\end{document}